# Harmful YouTube Video Detection:

# A Taxonomy of Online Harm and MLLMs as Alternative Annotators

Wonjeong Jo[1], Miki Wesołowska[2], and Magdalena Wojcieszak[1,2]

[1]Department of Communication, University of California, Davis

[2]Center for Excellence in Social Science, University of Warsaw

## Author Note

**Wonjeong Jo** https://orcid.org/0009-0006-7107-7349

**Magdalena Wojcieszak** https://orcid.org/0000-0001-5456-4483

Correspondence concerning this article should be addressed to Wonjeong Jo, 167 Kerr Hall, One Shields Avenue, Davis, CA 95616. Email: wjo@ucdavis.edu

## Abstract

Video platforms, used by billions globally, are criticized for exposing users to harmful content. Yet, detecting online harm remains challenging due to inconsistent approaches to what constitutes harm and limitations of human annotation. We address both challenges. Theoretically, we integrate and expand existing taxonomies and platform guidelines to multimodal contexts and develop a comprehensive taxonomy for online harm. We categorize online harm on video platforms into six content-based categories: Information, Hate and harassment, Addictive, Clickbait, Sexual, and Physical harms. Methodologically, we propose and systematically test multimodal large language models (MLLMs) as alternative annotators of harmful videos. We curate a dataset of 19,422 potentially harmful YouTube videos and annotate them using GPT-4-Turbo, non-trained crowdworkers, and trained annotators. We discuss the potential applications of MLLMs in platform content moderation and social science research, particularly in contexts where they perform comparably to human annotators in binary and multi-label classification tasks.

*Keywords*: multimodal analysis, online harm, social media, large language models, communication methods

**Harmful YouTube Video Detection:**

**A Taxonomy of Online Harm and MLLMs as Alternative Annotators**

5.07 billion people use social media platforms and this number has increased over the past decade (Statista, 2024). In this context, there are serious concerns that platforms direct users to harmful content. A user interested in herbology may encounter misinformation and a sad adolescent may easily watch content about depression or suicide (Hilbert et al., 2025). In fact, 66% of UK adults report encountering harmful content on social media (Turing Institute, 2023), audits find that self-harm and suicide-related posts on Instagram range from 9% to 66% (Picardo et al., 2020), and other work shows that YouTube often promotes drugs, alcohol, and bullying (Hattingh, 2021). Accordingly, 90% of US adults say that social media have harmful effects (Statista, 2023) and research finds that it leads to eating disorders and mental health issues, among other problems (Haidt & Twenge, 2023; Hartas, 2021).

Although platforms have community guidelines to regulate harm, content moderation presents two challenges. Theoretically, due to the subjective and context-dependent nature of harm, we lack a systematic and agreed upon understanding of harm. Platforms consider different content as harmful (see Arora et al., 2023) and researchers differently define and measure harm, as detailed below. In addition, extant operationalizations apply to text-based messages. Given the power of visuals in generating attention (Brosius, 1993) and effects (Powell et al., 2015) as well as the increasing popularity of video platforms, such as YouTube, Instagram, or TikTok (Faverio, 2025), it is essential to adapt and update existing taxonomies for multimodal context.

Methodologically, harm detection often involves human moderators (Roberts, 2016) and platforms' machine learning models and academic classifiers are trained with human labels.

However, given the mental toll on human annotators (Dang et al., 2018) as well as the resources for manual annotation, this approach is not scalable for large volumes of rapidly changing data.

This project addresses these issues. We first propose an overarching taxonomy for online harm. We synthesize existing taxonomies and platform guidelines and adapt to visual online environments. Our taxonomy contains six non-mutually exclusive categories: Information, Hate and harassment, Addictive, Clickbait, Sexual, and Physical harms. Each category is identifiable within a multimodal context, making it applicable to video platforms.

Next, we propose Multimodal Large Language Models (MLLMs) as alternatives to human annotators for identifying harm. In addition to conceptualizing harm, it is key to develop methods for identifying such content online. Recent work reveals remarkable capabilities of Large Language Models (LLMs) across text annotation tasks (e.g., Ostyakova et al., 2023), including in harmful text (e.g., Hoes et al., 2023), but their use for multimodal and multi-label content is underexplored (Kroon et al., 2023). We leverage GPT-4-Turbo and compare its performance to non-trained crowdworkers, setting trained annotators as a gold standard. We focus on YouTube videos,[1] yet our taxonomy and methodological approach can be applied to other platforms. We compile a large set of potentially harmful YouTube videos and use 19,422 videos for labeling by GPT-4-Turbo, crowdworkers, and trained annotators, as detailed below.

Our project offers several contributions. First, we integrate and extend past scholarship in communication and (computational) social science to develop an online harm taxonomy. This integrates separate scholarly efforts and can set a framework for operationalizing harm across platforms. Second, we address the increasingly prevalent multimodal and multi-label contexts in which online harm is created and consumed. We expand the application of MLLMs to binary tasks (classifying videos as harmful or harmless) and multi-label tasks (assigning one or more harm categories) for multimodal data. We show that MLLMs and crowdworkers have different strengths and limitations across classification tasks and evaluation conditions, and that MLLMs outperform crowdworkers on several performance metrics. We discuss how MLLMs' strengths in annotating potentially harmful multimodal content may support a broader range of social science research applications beyond platform moderation.

[1] YouTube is the most popular platform, used by 95% of teens (Vogels & Gelles-Watnick, 2023) and 81% of the American population (Auxier & Anderson, 2021).

This article is divided into two parts. First, we describe extant work on online harm taxonomies and show how we expand it to propose a comprehensive taxonomy for video platforms. Second, we outline the approaches used for harm detection, review recent work using LLMs for classification, and outline our data, classification methods, and results.

## Harmful Online Content

Harmful content is an expansive term, understood and measured differently across platforms and scholars. Researchers use distinct terms, such as *abuse* (Banko et al. 2020) or *problematic* (Yesilada & Lewandowsky, 2022) or *detrimental* (Gongane et al., 2022) content. Scholars also often focus on individual harms, such as self-harm or hate speech (Gongane et al., 2022; Hattingh, 2021), or on message features, such as incivility (Stoll et al., 2020), intolerant speech (Rossini, 2022), or toxicity (Google Jigsaw, n.d.; Lee et al., 2022). Taxonomies can minimize this fragmentation and facilitate joint understanding. Indeed, several efforts exist across academia, platforms, and non-profits (see Table 1). For instance, Banko et al. (2020) identify four harm types: hate and harassment, self-inflicted harm, ideological harm, and exploitation. Scheuerman et al. (2021) also categorize harms as physical, emotional, relational, and financial. In turn, Shelby et al. (2022) propose representational, allocative, quality-of-service, interpersonal, and social harms as their categories.

**Table 1**

*Existing Online Harm Taxonomy and Relevant Terminology*

| Type | Term | Categorization | Key source |
|---|---|---|---|
| Taxonomy | Harmful content | Hate and harassment, Self-inflicted harm, Ideological harm, Exploitation | Banko et al., (2020) |
| | Online harm | Physical harm, Emotional harm, Relational harm, Financial harm | Scheuerman et al. (2021) |
| | | Threats to personal and community safety, Harm to health and well-being, Hate and discrimination, Violation of dignity, Invasion of privacy, Deception and manipulation | Broom (2023) |
| | Sociotechnical harms | Representational harms, Allocative harms, Quality-of-service harms, Interpersonal harms, Social/societal harms | Shelby (2022) |
| Relevant concepts | Problematic content | Extreme right/left, Hate speech, Extremism, All-right, Islamophobia, Islamist extremism, Extremist messages, Mis/Disinformation, Conspiracy theories, Radicalization | Yesilada & Lewandowsky (2022) |
| | Inappropriate content | Fake news, Disinformation, Satire news, Rumor, Clickbait, Hate speech, Cyberbullying, Profanity, Toxic language, Abusive language, Sarcasm | Gongane et al. (2022) |
| | Toxicity | Severe toxicity, Obscene, Insult, Identity hate, Threat | Google Jigsaw (n.d.) |
| | Incivility | Disrespectful language such as name-calling, insulting language, and profanity | Coe, Kenski, & Rains (2014) |
| Video platform | Content moderation criteria | Spam and deceptive practice, Sensitive content, Violent or dangerous content, Regulated goods, Misinformation, Educational/Documentary/Scientific, and Artistic content | YouTube |
| | | Violence and criminal behavior, Safety, Objectionable content, Integrity and authenticity, Respecting intellectual property, Content-related requests and decisions | Instagram and Meta |
| | | Safety and civility, Mental and behavioral health, Sensitive and mature themes, Integrity and authenticity, Regulated goods and commercial activities, Youth safety and well-being | TikTok |

*Note.* See SM Table S2 for the full subcategory information under each categorization.

Although important, these taxonomies have a few limitations. First, specific harms are assigned to different categories in different taxonomies. For instance, information harms, which entail mis- or disinformation, are classified as ideological harms in Banko et al. (2020), as sociotechnical harms in Shelby et al. (2022), and placed separately in yet other frameworks (Hattingh, 2021; Shahi & Kana Tsoplefack, 2022). Each approach explains a fragment of online harm, but these different taxonomies integrated together can provide a more comprehensive insight. Second, these taxonomies conflate categories that are content-specific (e.g., insult, violence), intention-based (e.g., scams, misinformation), or effects-determined (e.g., emotional or relational harm). This is not only conceptually inconsistent but also makes harm difficult to measure. For example, Scheuerman et al.'s (2021) taxonomy is derived from user interactions, where harm is based on user perception rather than observable content features. Third, many categorizations are developed based on and for text only (e.g., Wikipedia, news, or YouTube comments). For example, toxicity, intolerance, and incivility rely on text-based linguistic characteristics, and the resulting classifiers cannot capture nonverbal information. Yet, given the rising share of video platforms, it is essential to scaffold a taxonomy that accounts for various harms, relies on observable indicators, and can apply to multimodal messages.

We address these gaps by integrating existing categorizations into an overarching framework. We draw on several definitions (Banko et al., 2020; Scheuerman et al, 2021; Shelby et al., 2022) to define online harm as content that is expected to cause offence or harm and which can be illegal (e.g., sexual exploitation or abuse), socially undesirable (e.g., drugs, violence), or cause physical or psychological distress (e.g., harassment, promoting self-harm). Furthermore, we apply this definition to multimodal contexts.

***Taxonomy for Online Harm in Video Platforms***

We outline our approach to creating a taxonomy for online harm in video platforms. To develop our taxonomy, we drew on the grounded theory approach (Glaser & Strauss, 1998), which is "a method of conducting qualitative research that focuses on creating conceptual frameworks or theories through building inductive analysis from the data" (Charmaz, 2006, p. 187). Rather than conducting open coding on raw video content, we began with a qualitative synthesis of existing taxonomies and platform guidelines of YouTube, TikTok, and Meta, treating these documents as our data source. Although extant taxonomies may be limited in scope or conceptual coherence, they offer structured categories informed by moderation practice and policy. By synthesizing and reorganizing their subcategories, we arrived at broader common categories. Our method thus sought to extend previous frameworks, identifying missing harm types, reconciling inconsistencies, and reorganizing categories to – again – make them detectable based on (multimodal) content alone. In SM (Supplemental Material) Table S1, we present these guidelines and show how our categorization aligns.

Our taxonomy is based on three principles. (1) *Each category is discernible within the multimodal context, encompassing text, audio, and images.* Categories not identifiable or observable within this context were excluded. (2) *A single piece of content can fall under multiple harm categories.* Each category may interact within the same or different modalities. For instance, a video narrating hate toward women while showing women being punched can be both hate and harassment (due to the audio) and physical harm (due to the visuals). (3) *Maximizing objective assessment.* Although the idea of harm naturally entails normativity and some subjectivity, we aim to minimize subjective judgment. We avoid value judgements, such as "problematic," "extreme," or "radical" content. Instead, each (sub)category is designed to be as verifiable as possible. Thus, we use categories that are agreed to be harmful based on extant taxonomies and platform guidelines, and on factors that are observable from the videos.

**Table 2**

*Taxonomy for Harmful Online Content on Video Platforms*

| Harm categories | Subcategories | Example cases |
|---|---|---|
| **Information harms** | Fake news | False rumors, fake celebrity deaths, non-factual information, health mis(dis)information |
| | Conspiracy theories | Chemtrails, flat Earth, moon landing, 9/11, political or vaccine conspiracies |
| | Unverified medical treatments | Vaccine denial/injuries/myths, unverified pill promotions, unverified herbal treatments |
| | Unproven scientific myths | AI-created scientific videos, nonscientific hypotheses (e.g., Pluto collided with Neptune, aliens in Area 51) |
| **Hate and harassment harms** | Insults and obscenities | Use of name-calling, insulting language, or profanity targeted at individuals or groups |
| | Identity attacks or misrepresentation | Misleading information or defamatory content about specific individuals |
| | Hate speech based on gender, race, ethnicity, age, religion, political ideology, disability, or sexual orientation | Violent incitement or hatred towards targeted people |
| **Addictive harms** | Online gameplay | Graphic game scenes, graphic game promotions |
| | Drug/smoking/alcohol promotion | Tutorials on growing or distributing drug plants, depictions of smoking or alcohol use, or glorification of drugs/smoking/alcohol |
| | Gambling-play videos | Recorded gambling videos, casino gameplay, advertisements for casinos or gambling services |
| **Clickbait harms** | Clickbait titles | Exaggerated, sensational, or hyperbolic titles, intending high click rates (*e.g., "You won't believe", "10 reasons you don't know," "Best/easiest hacks to"*) |
| | Get-rich-quick schemes or fishing financial hacks | Unverified financial hacks (e.g., Earn $800 in 5 minutes) or instructions on fishing/scam websites |
| | Gossip promotion | Sensational gossip, such as defamation videos, Hollywood rumors, and plastic surgery reveal |
| **Sexual harms** | Erotic scenes or images | Extracted clips from films, sexually explicit messages or images, or description of the actual use of sex toys |
| | Depictions of sexual acts and nudity | Pornography, masturbation, groping, upskirting, and depictions of sexual fetishes |
| | Sexual abuse | Videos depicting sexual coercion, or exploitation, including non-consensual acts |
| **Physical harms** | Self-injury and suicide | Videos alluding, advocating, or describing self-injury or suicide |
| | Eating disorder promotion | Promotion of eating disorders (e.g., pro-ana, pro-mia, thinspiration) |
| | Dangerous challenges and pranks | Depiction of dangerous behaviors that pose serious danger (e.g., tide pod challenge, blackout challenge) |
| | Violent graphic content | Fight or killing scenes, gun depictions, blood/dead body depictions, accident scenes, physical abuse of humans and animals |

We identify categories of online harm (see Table 2). First, **Information Harms** relate to the dissemination of false information that misleads and deceives people. This includes fake news, misinformation, disinformation, conspiracy theories, unverified medical treatments, and unproven scientific myths. Such content can undermine public trust in science, generate risks to public health, and exacerbate cynicism and extremism (Lazer et al., 2018). **Hate and harassment Harms** pertain to the promotion of hatred towards specific groups. This includes insults and obscenities, identity attacks, and hate speech based on gender, race, ethnicity, age, religion, political ideology, disability, or sexual orientation. Such content fosters hostility and discrimination with the intent to degrade the targeted groups. It may lead to psychological distress, (online and offline) demobilization of groups subjected to hate (Keum et al., 2024), and offline harassment and violence (Hefit & Ausserladscheider Jonas, 2020).

**Addictive Harms** refer to content that depicts or glorifies behaviors associated with addiction, including excessive gaming, gambling, or substance use (drugs, smoking, or alcohol). Such content can influence public perceptions, encourage compulsive engagement, and normalize restricted or unhealthy activities, potentially leading users to develop addictions or lose money (Moreno & Whitehill, 2014; Pienaar et al., 2017; Wakefield et al., 2003). **Clickbait Harms** are characterized by sensationalized content designed to attract clicks without delivering valuable information. This includes exaggerated headlines intended to boost click rates, unverified financial schemes[2], and sensational gossip or defamatory videos (Chakraborty et al., 2016; Chen et al., 2015). Clickbait misleads viewers, degrades the quality of the information ecosystem, and increases exposure to spam and phishing attacks (Zannettou et al., 2018). **Sexual**

[2] We include financial harms within this category because our taxonomy is organized around observable content-level mechanisms, and such schemes typically rely on the same deceptive, attention-driven strategies that characterize clickbait.

**Harms** encompass explicit sexual content, including erotic scenes, sexual acts and nudity, and sexual abuse, inappropriate for a general audience due to their sensitive or non-consensual nature. Such content can foster positive perceptions of sexual exploitation, reinforce misguided sexual roles (Ward et al., 2016), and lead to desensitization (Daneback et al., 2018). Finally, **Physical Harms** involve content portraying dangerous behaviors and graphic violence, e.g., self-injury, suicide, eating disorders, and dangerous challenges. Such content poses risks to viewers' health, can lead users to mimic the depicted activities (e.g., challenges leading to death; Kobilke & Markiewitz, 2021), and normalizes violence and self-inflicted harm (Khasawneh et al., 2020).

Although various research efforts have distinct foci, e.g., aiming to identify or mitigate specific harms, such as hate speech or misinformation, our taxonomy can expand these efforts. It systematically adapts existing categories and taxonomies to video platforms and may facilitate comparative research that estimates the prevalence of distinct harms across platforms. In addition, the taxonomy may facilitate harm mitigation by providing versatile categories that are platform agnostic and – as such – can be applied by policymakers to various platforms. While the evolving nature of social media content may introduce new types of harms, these – we suspect – can be accommodated within our framework.

## Identifying and Classifying Online Harm

It is essential to develop methods for identifying such content online. Traditionally, human annotation is seen as most accurate and often used in training and evaluating classifiers (Plank, 2022). Although this is a gold standard, manual labeling is not scalable for large datasets, particularly for videos, due to the time and costs needed for recruitment, training, and labeling (Shvetsova et al., 2023). The strain on human annotators who review harmful content is another key problem. Repeated exposure to harm has negative psychological and emotional effects (Dang et al., 2018), with Microsoft's moderators reporting post-traumatic stress disorder (Levin, 2017). Thus, alternative annotation methods that are cost-effective, faster, and devoid of ethical concerns are needed.

Against these challenges, recent studies show remarkable capabilities of LLMs, particularly OpenAI's ChatGPT, in this task. ChatGPT is cheaper and faster than crowdworkers, offering quality results that are comparable to, or even exceed, those of humans (Ostyakova et al., 2023). These studies propose LLMs as a viable alternative to humans for text annotation. For instance, ChatGPT outperforms crowdworkers and trained human annotators in classifying political messages (Törnberg, 2023) or detecting topics in online texts (Gilardi et al., 2023). Germane here, ChatGPT has strong capabilities in detecting misinformation, hate speech, and vaccine-related stances (Hoes et al., 2023; Huang et al., 2023; Kim et al., 2024).

Building on this work, the advent of multimodal LLMs (MLLMs) creates new possibilities for classifying multimodal data. Our study uses GPT-4-Turbo, one of OpenAI's MLLM APIs, and compares its performance in identifying harmful videos with that of untrained crowdworkers, compared against trained annotators' labels. Applying our taxonomy, we determine (a) whether a video is harmful, and (b) to which category or categories the content belongs, integrating both text and image in the process.[3]

***Data: Potentially Harmful YouTube Video Set***

We constructed a dataset of potentially harmful YouTube videos. Given that content that violates platform policies is often removed by platforms, leaving limited access to such content (Arora et al., 2023), we leveraged three approaches to obtain a large dataset representing all six harm categories. The details on each approach are available in SM 2.

---

[3] This study was approved by the IRB of the authors' institution. See SM 10 for ethical statements.

- Keyword-based approach. We developed a list of specific English keywords (n = 167) targeting each harm category from the taxonomy. This list incorporates keywords from previous studies and from platform community guidelines. We retrieved between 50 to 500 search results per keyword, applying recency (which collects videos sorted by the most recent) and relevance filters (which gather videos sorted by relevance, capturing the top results from queries).
- Channel-based approach**.** A research assistant compiled a list of 92 YouTube channels discussed as harmful in online communities such as Reddit (e.g., misinformation, clickbait financial hacks) or notorious for spreading harm (e.g., self-harm, drug encouragement, or sexual content).
- External dataset integration. We also relied on lists of problematic channels from published studies. We included channels identified as anti-feminist and White identitarian (Mamié et al., 2021; Ribeiro et al., 2020) as well as those associated with misinformation (Knuutila et al., 2021; Liaw et al., 2023; Papadamou et al., 2022).

Through these convergent steps, we identified 60,906 *potentially* harmful videos. For each, we retrieved text metadata (link, title, channel name, description, transcript, publish date, duration, views) and image data (14 image frames and a thumbnail). To collect text metadata, we utilized `YouTube Data API v3`[4] and `YouTubeTranscriptAPI`. For image extraction, we randomly selected 15 frame positions from the video using `yt_dlp` and `opencv-python (cv2)`. The frames were initially saved as PNG and later converted to base64 format to be compatible with the GPT-4-Turbo API. From the initial dataset, we sampled 19,422 videos for classification. Note that the collected videos are highly likely to be harmful, but not every video necessarily is.

[4] In terms of data use, the YouTube platform supports the CC BY license.

This approach also increases the number of boundary cases in the dataset. For example, not all videos retrieved using harmful keywords are themselves harmful (e.g., the keyword *"suicide"* may retrieve educational or prevention-related videos), and not all videos from channels known for harmful content are harmful (e.g., a drug-related channel may also upload ordinary videos).

### *Harmful Video Classification*

These videos were classified by: (a) GPT-4-Turbo, (b) untrained crowdworkers, and (c) trained annotators. Figure 1 provides an overview of the annotation process by three actors. All the steps outlined below are detailed in SM 3, 4, and 5.

**Figure 1**

*Classification Pipeline*

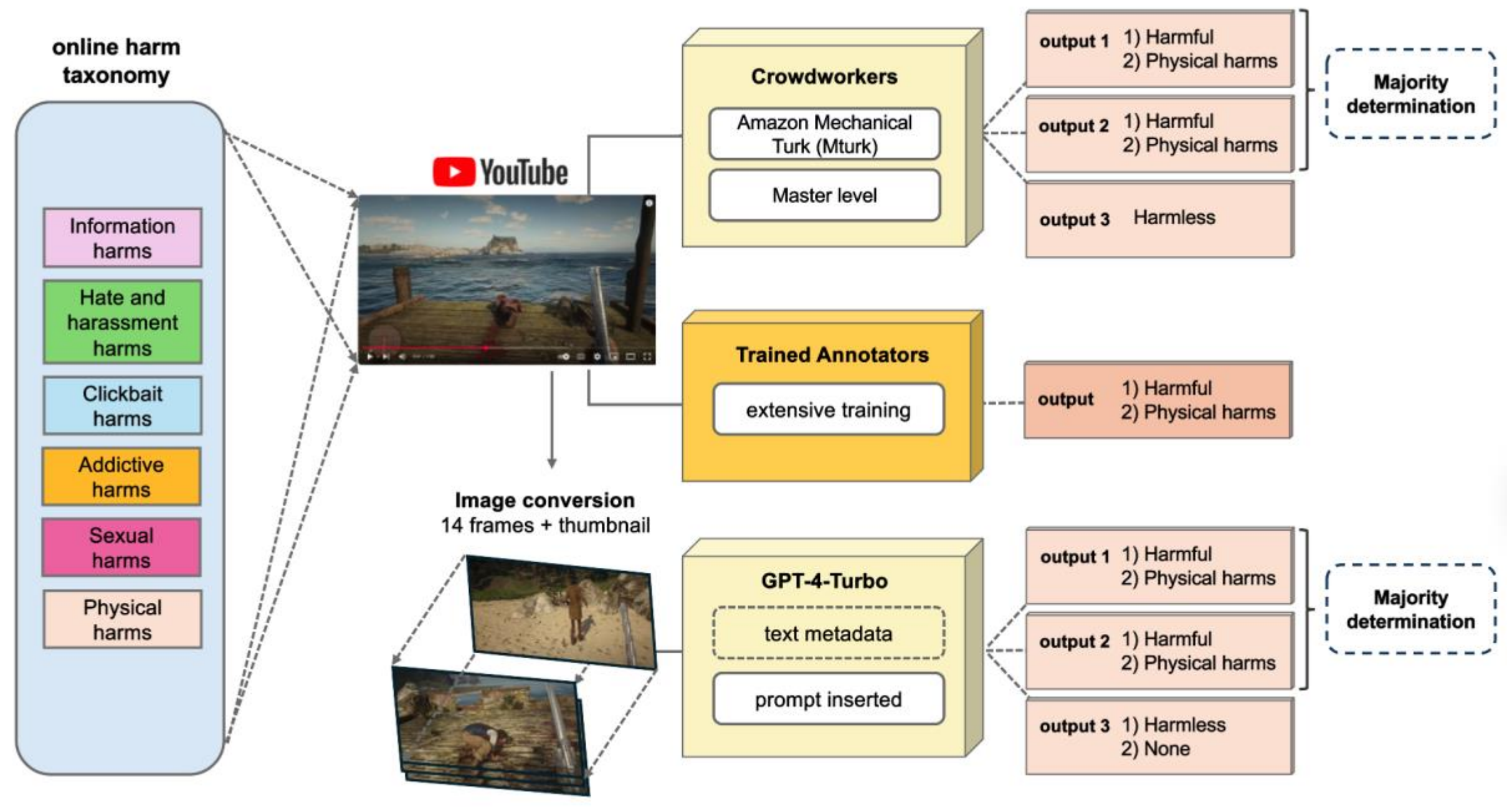

*Note*. The full GPT-4-Turbo prompt is provided in Table 3. The example video in the figure describes the use of marijuana and was therefore categorized as Addictive harms by trained annotators.

**Trained annotators.** We set the trained annotators' annotations as the gold standard. Ten undergraduate students majoring in communication underwent extensive training on the taxonomy and held periodic meetings to resolve any ambiguous cases and discuss disagreements. In classification, they were instructed to classify each video provided to them into one or more harm categories, or as harmless, unavailable (for removed videos), or unsure. When reviewing videos longer than 5 minutes, the trained annotators could skim by adjusting the mouse pointer or playback speed. The training materials and procedures are detailed in SM 3. In sum, the trained annotators labeled 19,422 videos over four months.

***Intercoder reliability.*** Intercoder reliability (ICR) was calculated using 11.4% (2,225) of the total videos. To calculate ICR, we assigned the same 265 videos to five pairs of trained annotators. Those within each pair watched the same videos, but different videos were assigned across groups. Additionally, 900 videos were assigned to random pairs. The average ICR was calculated for six groups: the five predetermined groups and one random group.

For binary classification (harmful vs. harmless), the percentage agreement was high at 88%. We report percentage rather than Cohen's Kappa, because the data are highly imbalanced (81% harmful videos). In such cases, Kappa can underestimate observed agreement by attributing much of it to chance agreement arising from the dominant category, even when actual agreement is high (Feinstein & Cicchetti, 1990; Holsti, 1969).[5]

For multi-label harm assignments, we assessed how consistently the two coders assigned the same harm categories to the same videos. We measured the overlap between the two trained annotators' labels within their multi-label assignments (six harm categories, harmless, unavailable). We also calculated ICR for each harm category. We calculated category-specific binary agreement by treating each harm category independently as a binary classification task

[5] When measured using Cohen's Kappa, it resulted in 0.37. The lower reliability of this metric reflects the imbalanced distribution of the sample dataset, as noted above. We additionally calculated Gwet's AC1, an inter-coder reliability coefficient for categorical data that adjusts for chance agreement but is less sensitive to category imbalance (AC1 = 0.72).

(present/absent). For example, if one annotator marked {Hate, Sexual} and another marked {Hate}, the agreement for the Hate and harassment category was coded as 1 (both marked it as present), but for the Sexual category was coded as 0 (disagreement on presence/absence). The highest agreement was for identifying Addictive harms (87.36%), followed by Sexual (78.00%), Clickbait (77.20%), Physical (71.80%), Information (71.10%), and Hate and harassment harms (58.80%).[6] See Figure 2 for a flowchart of the ICR calculation.

**Figure 2**

*Intercoder Reliability Calculation Procedure Between 10 Trained Annotators*

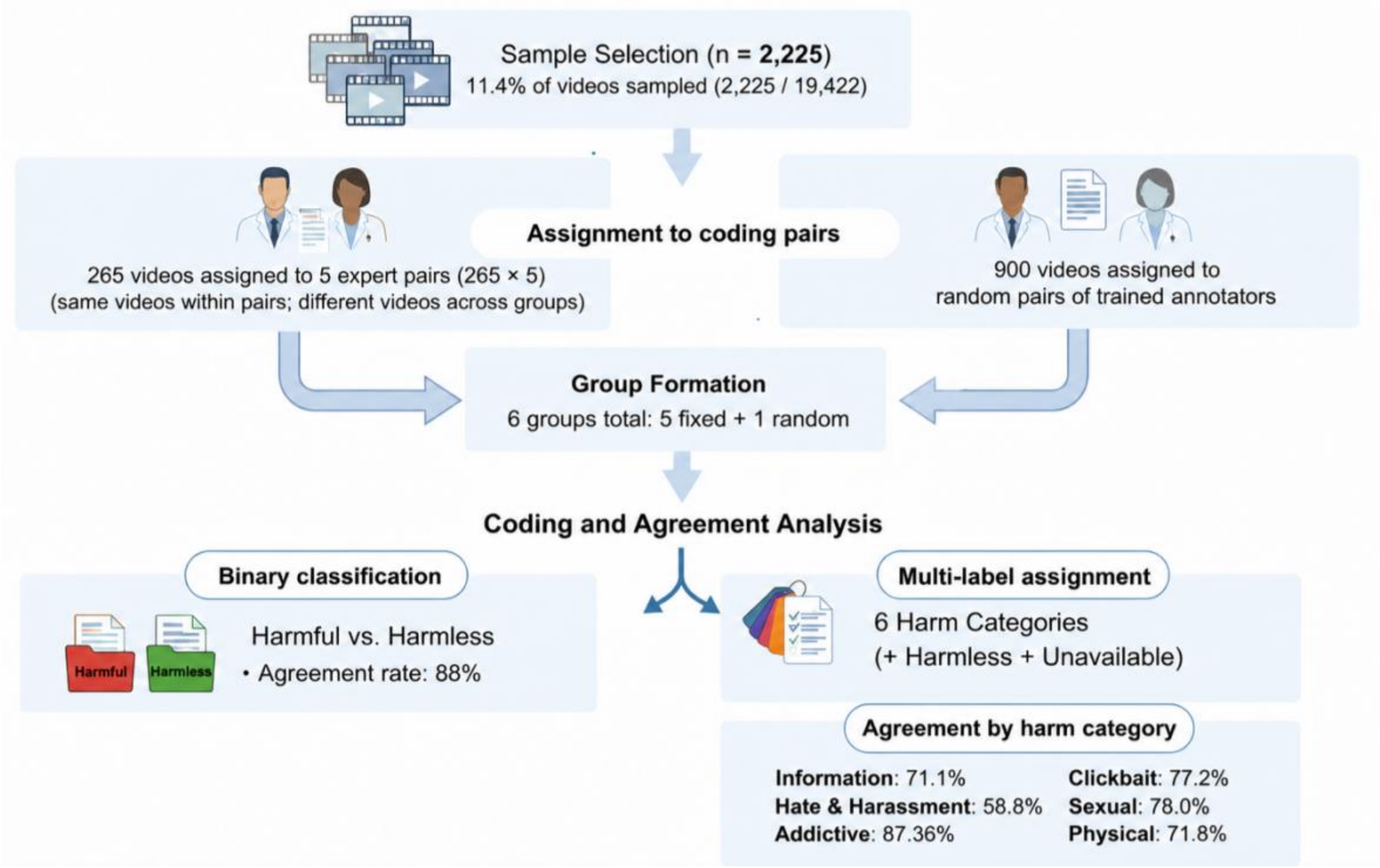

*Note*. Details on the training procedure for trained annotators are provided in SM 3.

---

[6] Lower agreement for the Hate and harassment category is consistent with prior literature, as such content is inherently context-dependent and multimodal, often requiring interpretation of both verbal and visual cues (Bui et al., 2025; Zhang et al., 2023). In contrast, categories with clearer visual signals (e.g., sexual harms) showed higher agreement. After calculating ICR, the trained annotators underwent additional rounds of extensive training to resolve remaining disagreements. We then recalculated ICR after this re-training, which resulted in 85.23% agreement. For comparison, the final agreement rates for the other categories were 96.44% for Addictive Harms, 95.39% for Sexual Harms, 91.38% for Information Harms, 89.69% for Physical Harms, and 81.20% for Clickbait. See SM 3 for the training materials.

**GPT-4-Turbo.** We used GPT-4-Turbo API.[7] We implemented prompt engineering before the annotation (see SM 4 for details) and finalized the prompt (see Table 3). The prompt was designed to be a zero-shot learning to avoid narrowing the classification reasoning to a limited set of examples. We structured our prompt into five parts: [Image Frames], [Task Disclaimer], [Coding Instruction], [Metadata], and [Question]. The process began by adding the image data (14 frames and 1 thumbnail), each converted to base64 format and resized to 768 pixels. Next, we placed [Task Disclaimer] and [Coding Instruction] and then inserted the text-formatted video metadata (i.e., video title, channel name, description, and transcript).[8] Lastly, we added the [Question] to request a parsable, formatted answer. For the API parameters, we used the "gpt-4-turbo" model with a temperature setting of 0.7, indicating moderate randomness which was a default setting when the study was conducted. We set a token limit of 50 for the output responses. To control for the model's randomness, we deployed three independent API calls and determined the majority answer. For example, if two API calls classified a video as harmful while the other one classified it as harmless, the video was regarded as harmful. If two or more APIs classified it as harmless, the video was classified as harmless.[9] We view multiple iterations as a useful baseline approach for LLM-based classification. Although deterministic settings (e.g., temperature = 0) settings may be easier to be used, outputs can still vary under those conditions (Atil et al., 2024; Ouyang et al., 2025). In addition, some recent LLM APIs

[7] Before selecting GPT-4-Turbo, we ran a pilot to compare OpenAI's ChatGPT, Meta's Llama, and Anthropic's Claude. At that time (Nov. 2023), the multimodal APIs of the other models were not stable enough to reliably process images, often returning random responses. This instability made it challenging to proceed with large-scale analysis.

[8] As our dataset contains some very long videos, we trimmed the transcripts to 3,000 words maximum (M = 1,177, Mdn.= 226, max.=7,077, min.= 0, Q3 = 1,913). For reference, a one-minute video with dense audio contains around 150 words. This maximized feasibility and was similar to human workers: trained annotators could skip parts of the video and crowdworkers similarly spent an average of 1.16 minutes per video.

[9] Robustness Test (pp. 19-21) confirms consistency between the deterministic (temperature = 0) and our iterative approaches.

(e.g., GPT 5) limit or remove control over the temperature parameter. As such, we consider our iterative approach a practical strategy across models and settings.

**Table 3**

*Final prompt for GPT-4-Turbo*

[Image frames] *14 image frames and 1 thumbnail image were provided here*

[Task disclaimer]
Your task is to classify YouTube videos as Harmful or Harmless, incorporating the image frames and metadata of a YouTube video. Please classify this video, making sure to reference the [Coding Instruction].

[Coding Instruction]
1. Information Harms (e.g., Fake news, Conspiracy theories, Unverified medical treatments, Unproven scientific myths)
2. Hate and harassment Harms (e.g., Insult, obscenities, Identity attacks or misrepresentation, Hate speech based on gender, race, religion, disability, sexual orientation)
3. Addictive Harms (e.g., Gameplay, Drug/smoking/alcohol promotion, Gambling videos or promotion)
4. Clickbait Harms (e.g., Clickbait titles - using exaggerated words, “You won’t believe,” “10 reasons you don’t know,” “Best/easiest hacks,” Get-rich-quick schemes and Fishing financial hacks, Gossip about celebrities)
5. Sexual Harms (e.g., Erotic scenes or images, Depictions of sexual acts and nudity, sexual abuse)
6. Physical Harms (e.g., Self-injury, Eating disorder promotion, Potentially dangerous challenges and pranks, Violent graphic content)

[Metadata] *title, channel information, description, and transcript were provided here*

[Question] Your answer should follow this format:
1) Harmful / Harmless
2) Corresponding one or multiple categories IF HARMFUL else None

**Crowdworkers.** We relied on MTurk (Amazon Mechanical Turk), a crowdsourcing platform. To maximize annotation quality, we recruited workers with a HIT approval rate above 95% and Master status.[10] We applied filters to recruit only English speakers aged 18 or older. As the task entailed harmful content, a warning message was displayed in the task title, and the task was published as private. Participants received $2 as compensation, and the median time taken to complete the task was 18.37 minutes (avg. = 25.75). 544 participants labeled 19,422 videos. The

[10] Before implementing the MTurk classification, we conducted a pilot test comparing the performance of master-level workers and non-verified lay workers using an identical set of 25 videos. Master-level workers spent more time on their tasks, had a higher filter question pass rate, and accurately re-wrote the harm taxonomy based on the provided instructions. Consequently, we employed only master-level workers to improve data quality.

annotators were demographically diverse. 60% male and 40% female, the majority Asian (60%) and white (35%), and 64.7% self-identified as heterosexual and 34.2% as homo- or bisexual (see Table S7).[11] This diversity is a strength, given that harm labeling can depend on demographic backgrounds (Kumar et al., 2021).

To ensure that all workers understood the classification taxonomy and to improve the response quality (Chmielewski & Kucker, 2020), we asked them to rewrite the six harm categories and complete a screening task in which they classified five videos as harmful or non-harmful. Workers who classified more than one video incorrectly were screened out, while those who passed proceeded to the main task (pass rate = 93.89%). In the main task, the workers classified 25 videos as harmful, harmless, or unavailable if the video was removed or unviewable. For harmful videos, the workers assigned one or more harm categories. As with GPT-4-Turbo classification, each video was labeled by three workers.[12]

***Results***

We assessed the performance of GPT-4-Turbo and crowdworkers in (a) binary classification (harmful vs. harmless), and (b) multi-label classification (one or more harm categories). Table 4 shows the binary classification results by each actor. Watching the same videos, trained annotators labeled the highest proportion as harmful (77.82%), while GPT-4-

---

[11] We speculate that the high prevalence of self-identified Asians and many non-homosexual workers may be due to either very unrepresentative nature of master MTurk workers or the fact that many participants may not have truthfully indicated their socio-demographics.

[12] The different classification procedures for trained annotators and crowdworkers were intentional. Each video was classified by a single trained annotator, whereas three different crowdworkers classified the same video and the majority response was used as the final label. The trained annotators received systematic training and repeated checks over an extended period to increase consistency. In contrast, concerns about the quality and reliability of crowdsourced annotations are well documented (e.g., Chmielewski & Kucker, 2020), and under a single-worker approach, an inattentive or random response could determine the final label. Therefore, we used three crowdworkers per video and majority voting as a quality-control measure. Cases in which no majority was reached among the three crowdworkers were excluded from the performance evaluation, as reported on pp.17-18.

Turbo labeled the lowest (54.04%). We computed ICR within the three GPT-4-Turbo API calls and the three crowdworkers separately on 30% of the dataset. Krippendorff's alpha among the three annotators yielded 0.78 for GPT-4-Turbo API calls and 0.21 for the crowdworkers.

**Table 4**

*Descriptive Information of Video Classification*

| | Trained annotator | Crowdworker | GPT-4-Turbo |
|---|---|---|---|
| Harmful | 15,115 (77.82%) | 12,668 (65.23%) | 10,495 (54.04%) |
| Harmless | 3,303 (17.00%) | 4,390 (22.60%) | 7,818 (40.25%) |
| Unavailable | 707 (3.64%) | 2,083 (10.72%) | 702 (3.61%)[a] |
| No agreement[b] | - | 220 (1.13%) | 227 (1.17%) |
| Removed[c] | 297 (1.53%) | 61 (0.31%) | 180 (0.93%) |

*Note.* [a] The unavailable case of GPT-4-Turbo includes classification failures (e.g., when GPT-4-Turbo fails to process images in the proper format).
[b] "No agreement" indicates cases where a majority decision was not made across three GPT-4-Turbo API calls or crowdworkers.
[c] From the "unavailable" labels, we separated the removal cases by automatically checking the HTML structure of each video. Since the videos were annotated at different times, the removal rates vary.

Figure 3-1 represents the distribution of harm categories by trained annotators, GPT-4-Turbo, and crowdworkers. We applied a majority selection to identify specific harm categories. Out of the harmful videos classified by crowdworkers, 4,976 (39.28%) had no majority harm categories, meaning that no category was selected by at least two of the three annotators. In contrast, GPT-4-Turbo had only 105 cases (1.00%) with no agreement out of harmfully classified videos, indicating greater within-method consistency than crowdworkers.

**Figure 3**

*Multi-Label. Harm Category Classification by GPT-4-Turbo, Crowdworkers, and Trained Annotators*

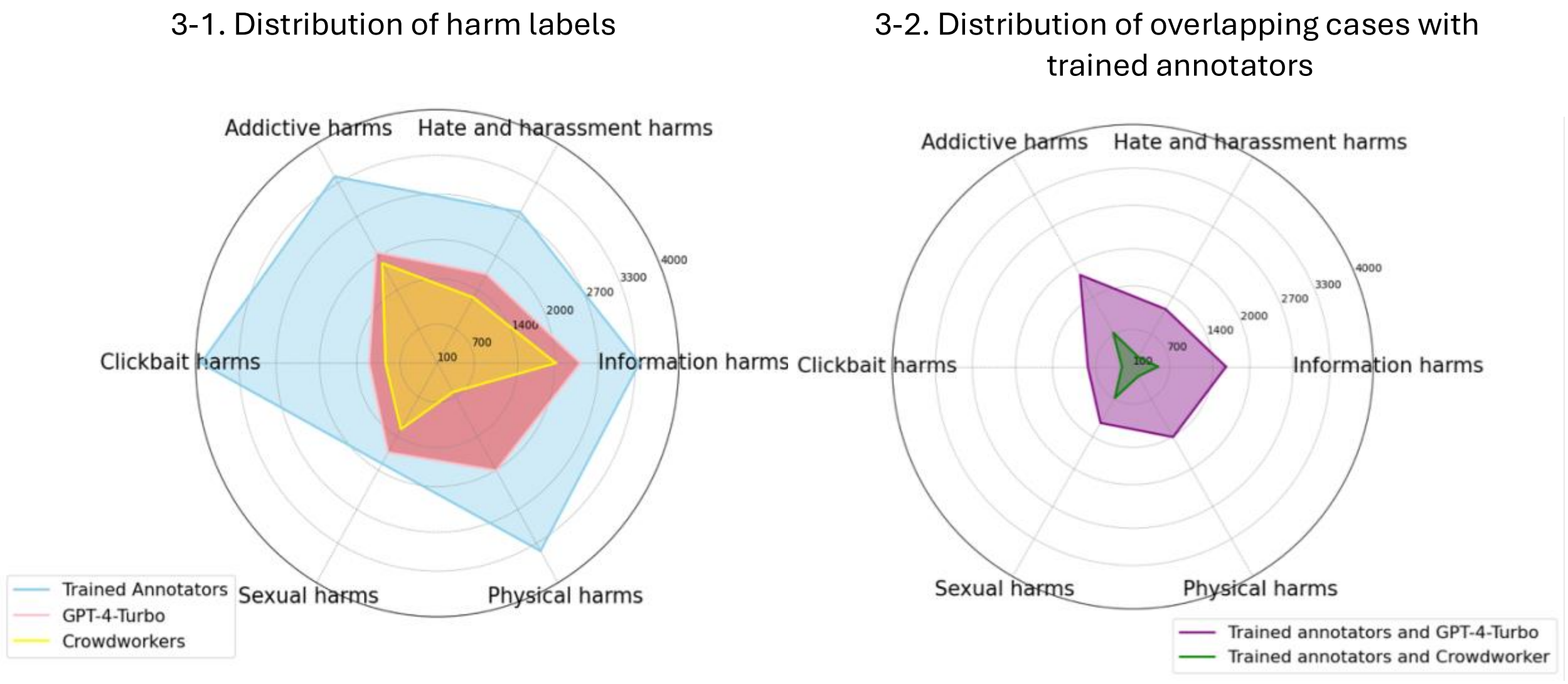

*Note*. Figure 3-1 presents the count of harm labels assigned by each actor. Figure 3-2 illustrates the count of overlapping cases between the initial labels from trained annotators and those from GPT-4-Turbo and crowdworkers. The complete label count is presented in Table S8.

**Performance evaluation for binary classification.** We compared the performance of GPT-4-Turbo and crowdworkers to trained annotators in the binary classification, excluding “unavailable,” “no agreement,” and “rejection” cases. The details are presented in Table 5. Both GPT-4-Turbo and crowdworkers similarly showed an accuracy of 66% compared against trained annotators. More specifically, GPT-4-Turbo showed superior performance in detecting harmless content with a higher specificity (true negative rate; 74.9%) compared to crowdworkers (29.0%). In contrast, sensitivity (true positive rate) was higher for crowdworkers (74.9%) than GPT-4-Turbo (64.1%), suggesting that crowdworkers are better at identifying harmful content. GPT-4-Turbo's lower sensitivity means it misses a larger share of harmful videos, despite the fact that harmful videos were the dominant category in the classification set. We report macro F1-score, precision, and recall as the most robust metrics for our unbalanced dataset where the majority are harmful videos.[13] On these aggregate metrics, GPT-4-Turbo outperformed crowdworkers.

**Table 5**

*Performance Comparison for Binary and Multi-label Classification*

| | **Binary classification** (Harmful vs. Harmless) | | **Multi-label classification** (Harm category annotation) | |
|---|---|---|---|---|
| | Crowdworker (n=16,199) | GPT-4-Turbo (n=17,536) | Crowdworker (n=13,071) | GPT-4-Turbo (n=14,487) |
| Accuracy | **0.660** | **0.660** | 0.136 | **0.390** |
| Sensitivity | **0.749** | 0.641 | 0.146 | **0.308** |
| Specificity | 0.290 | **0.749** | 0.891 | **0.941** |
| Macro F1 score | 0.514 | **0.595** | 0.260 | **0.579** |
| Macro Precision | 0.515 | **0.614** | 0.458 | **0.815** |
| Macro Recall | 0.519 | **0.695** | 0.195 | **0.462** |

*Note*. All values were evaluated against the trained annotators' labels. Ns differ because "unavailable," "no agreement," and "rejection" cases were excluded, and the multi-label comparison is restricted to videos labeled as harmful by trained annotators. Multi-label metrics were computed from binary six-bit strings indicating the presence (1) or absence (0) of each harm category, using the scikit-learn Python package (*sklearn*).

**Figure 4**

*F1-score, Precision, and Recall of GPT and Crowdworker across Harm Categories*

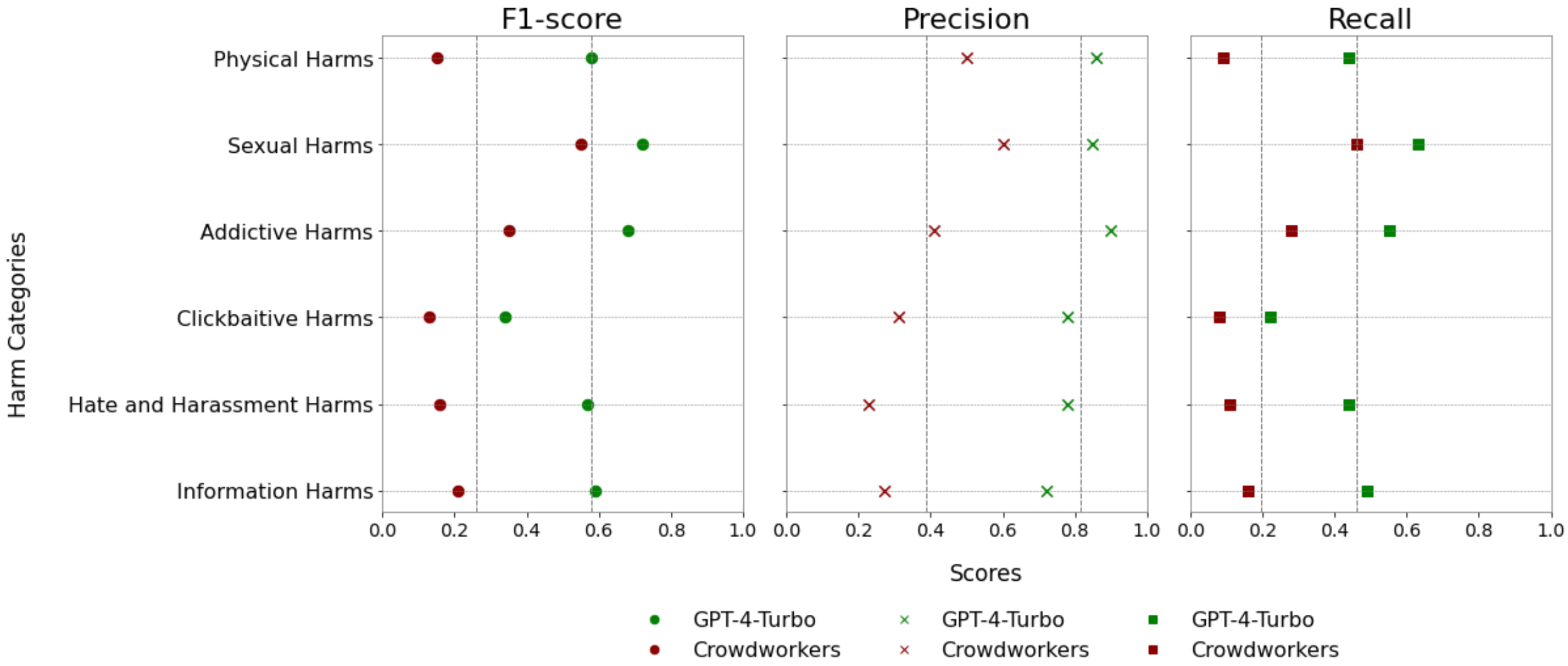

*Note*. Each metric is visualized in a separate plot. The dashed lines represent the macro F1-score, precision, and recall for GPT-4-Turbo and crowdworkers across the harm categories.

---

[13] Whereas the accuracy for binary classification can be inflated in imbalanced distributions, F1-scores provide a more reliable measure by penalizing extreme values.

**Performance evaluation for multi-label classification.** We also assess how well GPT-4-Turbo and crowdworkers classify harm into categories, focusing on videos labeled as harmful by trained annotators, excluding those marked as "unavailable" or with "no agreement" by GPT-4-Turbo API calls or crowdworkers.[14] We transformed the multi-label annotations into binary six-bit strings, with each bit indicating the presence (1) or absence (0) of a harm category. We then calculated F1-score, precision, and recall across the six categories. A total of 13,071 videos were used to compare trained annotators with crowdworkers, and 14,487 videos for comparing the annotators with GPT-4-Turbo. Figure 4 shows that GPT-4-Turbo outperforms crowdworkers in identifying all six harms. GPT-4-Turbo matched trained annotators' labels with an average (macro) F1 score of 57.9%, compared to crowdworkers' 26.0%. GPT-4-Turbo also achieved higher precision and recall than crowdworkers. However, both GPT-4-Turbo and crowdworkers showed relatively lower recall compared to precision, indicating that they tended to miss harm-category labels assigned by trained annotators. Accordingly, Figure 3-2 also supports a higher agreement between trained annotators and GPT-4-Turbo than between the annotators and crowdworkers. Looking at the individual harm categories, both GPT-4-Turbo and crowdworkers demonstrated the highest performance in detecting Sexual harms but the lowest performance in detecting Clickbait harms.

To further examine category-level agreement across the six harm categories, we compared where GPT-4-Turbo and crowdworkers disagreed with trained annotator labels. Cases in which crowdworkers disagreed with trained annotators, while GPT-4-Turbo did not, were most frequent for Information harms (22.7%) and Addictive harms (16.7%). These categories often require contextual interpretation of audio-visual cues. In contrast, Sexual harms showed the lowest overall disagreement rate, with 86.1% of videos agreed upon by all annotators, likely because these cases are more likely to involve visually salient signals. Clickbait harms showed the highest rate of mutual disagreement, where both GPT-4-Turbo and crowdworkers disagreed with trained annotators (14.8%).

---

[14] This resulted in a dataset where trained annotators' labels included one or more harm categories, while GPT-4-Turbo and crowdworker labels could include harm categories along with harmless label.

There are several possible reasons why crowdworkers performed worse on the multi-label task. First, their compensation may have been insufficient, limiting their motivation and engagement; performance might improve under different compensation settings.[15] Second, multi-label harm-category annotation requires annotators to understand and consistently apply a fine-grained codebook. Without the extensive training provided to trained annotators, crowdworkers may have been more likely to confuse adjacent or overlapping categories. Lastly, they likely varied in attention, cultural background, and interpretation of category boundaries, and such variation becomes especially consequential in a multi-label setting.

**Robustness Tests**

In addition to the main analysis, we conducted several robustness tests to confirm that our findings hold across different settings. All details on robustness tests are provided in SM 8.

**Deterministic approach in LLM setup**. Our main classification used three API call iterations with majority voting to mitigate randomness inherent in probabilistic models (default temperature = 0.7). To assess whether results are consistent under deterministic conditions, we additionally ran the model[16] at temperature = 0 on a subsample (n = 1,660) with balanced harmful and harmless videos based on original GPT-4-Turbo labels. Results were highly consistent with the original labels (81.00% agreement), indicating that our iterative approach produced results comparable to those obtained under a fixed temperature setting. Also, when these newly classified labels were evaluated against trained annotator labels, the accuracy was 74.3%, outperforming crowdworker performance (66.5%) and further supporting our main finding. The LLM also showed higher precision (0.85 vs. 0.78), sensitivity (0.81 vs. 0.79), specificity (0.49 vs. 0.22), and F1 score (0.83 vs. 0.79) than crowdworkers.

[15] In setting compensation, we followed prior work on MTurk-based annotation tasks (e.g., Kumar et al., 2021). In addition, for the robustness test in which 800 balanced videos were classified by both MLLM and crowdworkers, we increased task compensation to $6.

[16] Since GPT-4-Turbo has been deprecated in favor of GPT-4o, we used GPT-4o as its direct successor.

**Newer models.** We also note that model performance may evolve or plateau over time. To avoid limiting findings to GPT-4-Turbo, we evaluated more recent models (i.e., OpenAI's GPT-5-mini and Google's Gemini-3-flash-lite) on the same 1,660 subsample. Both models showed higher accuracy than crowdworker labels in binary (GPT-5-mini: accuracy = 68.51%, F1 = 0.78; Gemini-3-flash-lite: accuracy = 71.48%, F1 = 0.80; crowdworkers: accuracy = 66.5%, F1 = 0.79) and multi-label classification (Macro F1 GPT-5-mini: 0.62, Gemini-3-flash-lite: 0.63; crodworkers: 0.22), consistent with our main findings. Notably, both newer models achieved higher accuracy than GPT-4-Turbo in the original analysis. This may indicate that our main findings based on GPT-4-Turbo may "underestimate" MLLM performance, as newer models may offer improved classification performance across tasks. At the same time, also consistent with our main finding, sensitivity was higher for crowdworkers (0.79) than both of the newer models (GPT-5-mini: 0.71; Gemini-3-flash-lite: 0.74), while specificity was far higher for the models (GPT-5-mini: 0.57; Gemini-3-flash-lite: 0.63) than crowdworkers (0.22). These replicated results overall suggest that our methodological framework can be readily applied to future models as they emerge, and that our findings should be interpreted as evidence of both the promise and current limitations of MLLMs rather than as a definitive benchmark tied to a single model version.

**Classification with a balanced sample of harmful and harmless videos.** Our dataset was deliberately enriched for harmful videos across six harm categories. (77.82%),[17] which raises the concern that classifiers may inflate accuracy by gravitating toward the majority class (Stoll, 2023). Additionally, harmless videos collected may not represent explicitly "benign" content, as they were collected using harmful keywords, from channels associated with harmful content, and from externally annotated datasets. To address this, we first constructed a balanced subsample (n = 800) with equal numbers of harmful and harmless videos. 400 harmless videos were collected using benign keywords (e.g., "tutorial," "parent playing with kids," "python," "nature walk") and manually reviewed by the authors. Harmful videos were randomly sampled from the original dataset labeled by trained annotators. Both MLLM[16] and crowdworker classifications were re-run on this subset. Results were consistent with the main findings. The MLLM achieved higher accuracy than crowdworkers in both binary classification (83.7% vs. 80.4%) and multilabel classification (48.3% vs. 40.1%). Notably, accuracy was higher than in the main analysis, likely because the use of benign keywords produced more clearly harmless videos, reducing the edge cases prevalent in the original dataset. Yet, its F1 score was slightly lower for binary classification (0.73 vs. 0.74) but higher for multilabel classification (0.52 vs. 0.44). Also similar to the main finding, sensitivity was higher for crowdworkers (0.73 vs 0.60) while specificity was higher for GPT-4o (0.98 vs. 0.85). These results indicate that our main findings are not substantially affected by class imbalance.

[17] This is because we could not determine whether videos were harmful prior to annotation; instead, the harmful versus non-harmful classification was the outcome of the annotation process. Explicitly harmful content is also relatively rare and difficult to identify, either because platforms moderate it or because harm is embedded within otherwise benign videos, such as in subtle harassment or stereotyping. For this reason, we adopted a diagnostic sampling strategy that enriched the sample for potentially harmful content by drawing videos from a list of potentially harmful YouTube channels. This approach ensured sufficient coverage and diversity across all six harm categories, which was essential for evaluating multi-label classification performance.

**Baseline comparisons.** We also conducted baseline comparisons to contextualize both GPT-4-Turbo and crowdworker performance. We evaluated their results against three reference baselines using the imbalanced main dataset (n = 19,422): a majority-class baseline, a random baseline, and a stratified random baseline. These baselines achieved accuracies of 0.81, 0.69, and 0.70 and F1 scores of 0.90, 0.81, and 0.82, respectively. GPT-4-Turbo and crowdworkers achieved lower accuracies (0.66 and 0.64) and F1 scores (0.74 and 0.78). However, the baseline results largely reflected the predominance of harmful videos in the dataset. In particular, the majority-class baseline achieved high accuracy and F1 by classifying every video as harmful but failed to identify any harmless videos (specificity = 0.00). In contrast, GPT-4-Turbo achieved higher precision (0.92) and specificity (0.75) than all three baselines. For multilabel classification, GPT-4-Turbo achieved a macro F1 score of 0.56, outperforming the majority-class baseline (0.30), the random and the stratified random baselines (0.26), and crowdworkers (0.36). GPT-4-Turbo also achieved the highest category-specific F1 score in five of the six harm categories, with Clickbait as the only exception. These results together suggest that GPT-4-Turbo's performance is not an artifact of class imbalance or majority-class prediction but remains robust across different evaluation settings.

## Discussion and Conclusion

Online harm is not only prevalent (Statista, 2023) but also can have detrimental effects (Haidt & Twenge, 2023; Hartas, 2021). Despite efforts by platforms and scholars, harm detection remains challenging. Theoretically, this is due to a lack of an overarching taxonomy of online harm, which minimizes its context-dependent and often user-perception-based nature. Methodologically, approaches to identifying harm largely rely on human moderators or classifiers trained on human-labeled data, which present ethical and scalability issues. We address these issues and offer both theoretical and methodological contributions.

Theoretically, we develop a comprehensive taxonomy for online harm by integrating existing taxonomies and platform guidelines. We, in particular, consider the power of visuals and the growing popularity of video platforms (Faverio, 2025). As visuals are often underexplored in existing harm definitions and operationalizations, our taxonomy is explicitly adapted to multimodal harms that are identifiable based on content alone. Our taxonomy proposes six categories: Information, Hate and harassment, Clickbait, Addictive, Sexual, and Physical harms, contributing to harm mitigation through its applicability to most types of social media and video platforms and facilitates their measurement.

Methodologically, we propose an approach to identify the specific harms leveraging an MLLM. We curate a large dataset of potentially harmful videos (n = 19,422) and have them labeled by GPT-4-Turbo, crowdworkers, and trained annotators. We compare their performance in binary classification (harmful vs. harmless) as well as multi-label classification (harm category assignment), compared against trained annotators as the gold standard.

We find that GPT-4-Turbo showed comparable or higher accuracy in both binary and multi-label classifications for multimodal data, a finding supported across multiple conditions, including (1) a deterministic setting with temperature = 0, (2) newer MLLMs, (3) a balanced sample with equal numbers of harmful and harmless videos, and (4) several baseline comparisons. At the same time, we identify important differences in the strengths of MLLMs and crowdworkers. In binary classification, MLLMs' superior performance is driven mostly by their ability to correctly identify harmless content, resulting in substantially higher specificity. By contrast, crowdworkers are more effective at identifying harmful content, achieving higher sensitivity.

In the main analysis, the performance gap between GPT-4-Turbo and crowdworkers was modest particularly in binary classification and both of them correctly predicted only 66% of trained annotator labels. Yet importantly, our robustness tests using newer models and a more balanced set of clearly benign and harmful videos show substantially smaller accuracy gaps in the binary classifications between MLLMs and trained annotators. We also note that, given the asymmetric classification process, in which MLLMs received only 15 images while human annotators had access to the full videos, the performance we report in the main analysis may "underestimate" the capabilities of future MLLMs.

Although our findings may suggest limited evidence of MLLMs, in their current form, for deterministic and independent tools for harm classification "in the wild" in platform moderation, we see important potential for their use in multimodal harm identification. For practical applications, MLLM annotations can support research workflows involving large, purposively collected datasets. Possible applications include proposing preliminary harm-category labels or harm flags for subsequent human verification, assisting researchers in applying a complex codebook more consistently, and identifying difficult or ambiguous cases for closer review. This approach may also support audits of recommender systems by helping researchers pre-organize and code videos collected along recommendation pathways, where the distribution of harmful and harmless videos may vary depending on the recommendation setting.

For more methodological applications, given the growing number of available MLLMs, researchers could also incorporate multiple models and use an ensemble approach, such as taking the majority label across models, to improve consistency and strengthen performance as an automated classification tool. We also encourage future research on human-in-the-loop systems, in which trained annotators provide examples, review uncertain cases, and correct model outputs. Such human feedback could be used to refine prompts or fine-tune models, thereby improving MLLM performance over time.

Despite our theoretical and methodological contributions to online harm mitigation, we leave a few areas open for future research. First, the taxonomy is not free from normative complexities. Although it is a strength that our categories are detected using multimodal content alone, without considering contextual factors or target audiences, it could be argued that "harm" is a joint function of the content and its recipients. That is, the same content can be seen as harmful by some individuals but harmless by others. We acknowledge this complexity, yet (scalable) harm detection systems need to assess harm at content level, as considering audience-specific perceptions. As such, our taxonomy not only integrates past work toward minimizing subjectivity and accounting for multimodal content, but also could be readily applied to harm identification at scale.

Second, we analyzed solely YouTube data. Expanding the analysis to TikTok or Instagram would offer a more comprehensive comparison across social media platform. We also note that we did not have access to the removed YouTube videos. To the extent that some content was removed precisely because it was deemed harmful, our final sample does not contain videos which could send the clearest signal (whether to human labelers or the MLLMs) about a certain harm. Because our dataset contains videos that are less explicit and – as such – potentially more challenging to classify than those that we could not access, the detected accuracy could have been higher for more explicitly harmful videos.

Third, ChatGPT has limited replicability due to its black-box training process. Also, as the multimodal capabilities of MLLMs continue to evolve, research should replicate our approach using other newer models. Indeed, our robustness tests suggest that newer models may achieve even stronger performance. We do not advocate for sole reliance on a specific model. Rather, we view MLLMs as promising tools that can assist human annotators—or, in some contexts, serve as alternatives to them—in multimodal classification tasks.

Lastly, although our findings suggest that MLLMs have potential as an automated assistant tool for harm classification, this conclusion should be interpreted with caution. For one, MLLMs performed well in distinguishing harmless from harmful videos, but their performance was lower for multi-label harm category assignment. For example, performance was notably weaker for Clickbait harms. These results may indicate that MLLMs may be especially useful for binary harm classification and for certain categories (e.g., Information harms), but their use for other harm categories requires caution and, ideally, continued expert validation. For the other, GPT-4-Turbo was better at identifying harmless than harmful videos. This is a critical limitation because real-world harm detection may require the opposite. Given the different strengths and limitations of MLLMs and human annotators, a human-in-the-loop approach may be especially valuable. For example, MLLMs could be used to pre-screen or organize large datasets, while trained human annotators review uncertain cases and verify potentially harmful content. Future work should continue to track model performance across harm categories and different models.

In sum, this study advances theorizing and methods on online harm. We offer a systematic taxonomy and an applicable methodology, which – we hope – can facilitate the ever more pressing work by scholars, industry, and organizations to measure and identify harms across social media platforms. We hope our work can serve as a foundation for future efforts and advance rigorous computational approaches to identify and mitigate online harm on platforms and in broader computational social science.

## Data Availability Statement

All classification outcomes and analysis data are available in https://zenodo.org/records/14647452.

## Disclose Statement

The authors report there are no competing interests to declare.